\documentstyle[12pt]{article}
\def\beq{\begin{equation}}
\def\eeq{\end{equation}}

\begin{document}
\begin{titlepage}
\begin{flushright}
BA-99-01 \\
April 6, 1999 \\
\end{flushright}

\begin{center}
{\Large\bf An  Improved Supersymmetric SU(5)
\footnote{Supported in part by  DOE under Grant No. DE-FG02-91ER40626
and by NATO, contract \\
~~~~~~~~~~~~number CRG-970149.}
}
\end{center}
\vspace{0.4cm}
\begin{center}
{\large Qaisar Shafi$^{a}$\footnote {E-mail address:
shafi@bartol.udel.edu} {}~and
{}~Zurab Tavartkiladze$^{b}$\footnote {E-mail address:
z\_tavart@osgf.ge} }
\vspace{0.5cm}

$^a${\em Bartol Research Institute, University of Delaware,
Newark, DE 19716, USA \\

$^b$ Institute of Physics, Georgian Academy of Sciences,
380077 Tbilisi, Georgia}\\
\end{center}

\vspace{1.5cm}

\begin{abstract}
By supplementing minimal supersymmetric $SU(5)$ (MSSU(5)) with
a flavor ${\cal U}(1)$
symmetry and two pairs of $\overline{15}+15$ `matter' supermultiplets, we
present an
improved model which explains the charged
fermion mass hierarchies and the magnitudes of the
CKM matrix elements, while avoiding the undesirable asymptotic mass
relations  $m_s=m_{\mu }$, $\frac{m_d}{m_s}=\frac{m_e}{m_{\mu }}$.
The strong coupling $\alpha_s (M_Z)$ is
predicted to be approximately $0.115$, and the proton lifetime is
estimated to be about five times
larger than the MSSU(5) value. The atmospheric and solar neutrino
puzzles are respectively resolved via maximal $\nu_{\mu }-\nu_{\tau }$
and small mixing angle
$\nu_e-\nu_s$ MSW oscillations, where $\nu_s$ denotes a sterile
neutrino. The
${\cal U}(1)$ symmetry ensures not only a light $\nu_s$ but also
automatic `matter' parity.

\end{abstract}

\end{titlepage}

\section{Introduction}

It is well known that MSSU(5) \cite{dim} provides
no explanation of the charged fermion mass hierarchies and mixings,
predicts the undesirable asymptotic relations
$m_s=m_{\mu }$,
$\frac{m_d}{m_s}=\frac{m_e}{m_{\mu }}$, and
cannot simultaneously account for the atmospheric and solar neutrino data.
In addition, taking account of supersymmetric and heavy threshold
corrections, the MSSU(5) value for the strong coupling is $0.126$
\cite{lan},
to be compared with the world average value of $0.117\pm 0.005$ \cite{dat}

In a recent paper \cite{bimax} we considered
an $SU(5)$ model supplemented by two key ingredients.  One is a
${\cal U}(1)$ flavor symmetry \cite{fro}, suitably implemented for
explaining
the
charged fermion mass hierarchies and mixings, and consistent with a
variety
of neutrino oscillation scenarios.  For instance, it was shown how
bi-maximal neutrino mixings could be realized for explaining the
atmospheric and solar neutrino data.  A second key ingredient
is the introduction of two pairs of vector-like `matter' superfields
belonging
to the  $\overline{15}+15$ representations of $SU(5)$.  They play an
essential role in avoiding the
undesirable asymptotic mass relations mentioned above
\footnote{The extended `matter' sector of $SU(5)$,
with additional $\overline{15}+15$ supermultiplets, leads to a
scenario quite different from the case which includes scalar $45$-plets
\cite{GJ}.}.

The purpose of this paper is to explore some key phenomenological
consequences of such an extended
$SU(5)$ scheme.  In particular, it turns out that the
$\overline{15}+15$ superfields play an essential role in reducing the
predicted strong coupling to $\simeq 0.115$, which is in excellent
agreement
with experiments.  Furthermore, they also have an
impact, albeit a modest one, on the proton lifetime.  It
turns out to be about five times longer than the MSSU(5)
value.

For obtaining a natural understanding of the charged fermion
mass hierarchies and magnitudes of the CKM matrix elements, we
supplement the ${\cal U}(1)$ flavor symmetry with a $Z_2$
${\cal R}$-symmetry.  The
latter helps in the generation of desired mass scales.
The resolution of
the atmospheric and solar neutrino puzzles necessitates in this
approach the introduction of a sterile neutrino state $\nu_s$ which,
thanks to the ${\cal U}(1)$ symmetry, can be kept light.
Maximal $\nu_{\mu }-\nu_{\tau }$
oscillations resolve the atmospheric neutrino anomaly, while
the small mixing angle $\nu_e-\nu_s$ MSW oscillations can explain
the solar neutrino data.  It turns out that the ${\cal U}(1)$ symmetry
also implies an automatic
$Z_2$ `matter' parity (including higher order terms).

\section{Extended Supersymmetric $SU(5)$: \\
Charged Fermion Masses and Mixings}

The scalar sector of $SU(5)$, which we consider here, in addition
to $\Sigma (24)$, $\bar H(\bar 5)$, $H(5)$ 
multiplets, also contains
$S$ and $X$ singlets. We introduce the symmetry
$Z_2\times {\cal U}(1)$, where $Z_2$ is an ${\cal R}$-symmetry.
Under $Z_2$,

$$
(\Sigma ,~\overline{H} ,~H ,~S)\rightarrow - 
(\Sigma ,~\overline{H} ,~H,~S)~,
$$
\beq
X\rightarrow X~,~~~~W \rightarrow - W~.
\label{z2}
\eeq
As will be discussed in more detail below, the
anomalous ${\cal U}(1)$ flavor symmetry is crucial for obtaining the
hierarchies among
fermion masses and mixings. The ${\cal U}(1)$ charges of the
`scalars'
are:

$$
Q_X=1~,~~~~Q_{\bar H}=-Q_H=2r~,
$$
\beq
Q_{\Sigma }=Q_S=0~,
\label{qsc}
\eeq
($r$ is undetermined for the time being).

The most general renormalizable scalar superpotential  
allowed under the symmetries  reads:

$$
W_S=-\Lambda^2S+\frac{\lambda }{3}S^3+\frac{h}{2}S {\rm Tr}\Sigma^2
+\frac{\sigma }{3} {\rm Tr}\Sigma^3+
$$
\beq
\overline{H}(\lambda_1S+\lambda_2\Sigma )H~,
\label{ws}
\eeq
where $\lambda $, $h$, $\sigma $ and $\lambda_{1,2}$ are dimensionless
couplings, and $\Lambda $ is a mass scale of order $M_{GUT}\equiv M_G$.
From (\ref{ws}), with supersymmetry unbroken, one obtains a non-vanishing 
$\langle \Sigma \rangle $ (and also $\langle S\rangle $) in the desirable
direction

\beq
\langle \Sigma \rangle ={\rm Diag }(2,~2,~2~,-3~,-3)\cdot V~,
\label{dir}
\eeq
with
\beq
V=\frac{h\Lambda }{(15h^3+\lambda\sigma^2)^{1/2}}~,~~~~
\langle S \rangle =\frac{\sigma \Lambda }{(15h^3+\lambda\sigma^2)^{1/2}}~.
\label{vevs}
\eeq
From (\ref{vevs}), assuming that $\Lambda \sim 10^{16}$~GeV, with all
coupling constants
of
order unity, we have

\beq
\frac{V}{M_P}\sim \frac{\langle S \rangle}{M_P}\equiv \epsilon_G
\simeq 10^{-2}~.
\label{epsG}
\eeq

As for the flavor ${\cal U}(1)$ symmetry, it is natural to consider it as
an anomalous gauge symmetry. It is well known that anomalous $U(1)$
factors can appear in effective field theories from strings.
The cancellation of its anomalies occurs through the Green-Schwarz
mechanism
\cite{gsh}. Due to the anomaly, the Fayet-Iliopoulos term

\beq
\xi \int d^4\theta V_A
\label{fi}
\eeq
is always generated \cite{fi}, where, in string theory, $\xi $ is given by
\cite{xi}

\begin{equation} 
\xi =\frac{g_A^2M_P^2}{192\pi^2}{\rm Tr}Q~.
\label{xi}
\end{equation}  
The $D_A$-term will have the form
\begin{equation}
\frac{g_A^2}{8}D_A^2=\frac{g_A^2}{8}
\left(\Sigma Q_a|\varphi_a |^2+\xi \right)^2~,
\label{da}
\end{equation}
where $Q_a$ is the `anomalous' charge of $\varphi_a $ superfield.

In ref. \cite{gia} the anomalous ${\cal U}(1)$ symmetry was considered as
a
mediator of SUSY breaking, while in ref. \cite{u1}, the anomalous Abelian
symmetries were exploited as flavor symmetries for a natural understanding
of hierarchies of fermion masses and mixings.

In our $SU(5)$ model, assuming ${\rm Tr}Q<0$ ($\xi <0$) and taking into
account (\ref{qsc}), we can ensure that the cancellation of (\ref{da})
fixes the VEV of $X$ field as:

\beq
\langle X\rangle =\sqrt{-\xi }~.
\label{vevx}
\eeq
Further, we will assume that

\beq
\frac{\langle X\rangle }{M_P}\equiv \epsilon \simeq 0.2~.
\label{epsx}
\eeq
The parameter $\epsilon $ is an important expansion parameter for
understanding the magnitudes of
fermion masses and mixings.

Together with the $(10+\bar 5)_i$ ($i=1, 2, 3$ is a family index)
matter multiplets, we consider two pairs 
$(\overline{15}+15)_{1,2}$ of `matter', which will play an important role
for
obtaining acceptable pattern of fermion masses.
The transformation properties of `matter' superfields under 
${\cal U}(1)$ are given in Table (\ref{t:fer}).
The relevant couplings will be\footnote{We assume that $Z_2$ ${\cal R}$
symmetry does not act on the matter superfields.}:

\begin{equation}
\begin{array}{ccc}
 & {\begin{array}{ccc}
\hspace{-5mm}~\bar 5_1~& \,\,\bar 5_2~~ & \,\,\bar 5_3~~~

\end{array}}\\ \vspace{2mm}
\begin{array}{c}
10_1 \\ 10_2 \\ 10_3
 \end{array}\!\!\!\!\! &{\left(\begin{array}{ccc}
\,\,\epsilon^5 ~~ &\,\,\epsilon^4~~ &
\,\,\epsilon^3
\\
\,\,\epsilon^3 ~~  &\,\,\epsilon^2~~ &
\,\,\epsilon
 \\
\,\,\epsilon^2~~ &\,\,\epsilon ~~ &\,\,1
\end{array}\right)\overline{H}\epsilon^a }~,
\end{array}  \!\!  ~~~~~
\label{10-5}  
\end{equation}

\begin{equation}
\begin{array}{ccc}
 & {\begin{array}{ccc}
\hspace{-5mm}~10_1& \,\,10_2 & \,\,10_3

\end{array}}\\ \vspace{2mm}
\begin{array}{c}
10_1 \\ 10_2 \\ 10_3
 \end{array}\!\!\!\!\! &{\left(\begin{array}{ccc}
\,\,\epsilon^6 ~~ &\,\,\epsilon^4~~ &
\,\,\epsilon^3
\\
\,\,\epsilon^4 ~~  &\,\,\epsilon^2~~ &
\,\,\epsilon
 \\
\,\,\epsilon^3~~ &\,\,\epsilon~~ &\,\,1
\end{array}\right)H }~,
\end{array}  \!\!  ~~~~~
\label{10-10}  
\end{equation}

\begin{equation}
\begin{array}{cc}
 & {\begin{array}{ccc}
\bar 5_1~~&\,\,\bar 5_2~~&\,\,\bar 5_3~~~~~~
\end{array}}\\ \vspace{2mm}
\begin{array}{c}
15_1\\ 15_2

\end{array}\!\!\!\!\! &{\left(\begin{array}{ccc}
\,\, \epsilon^5~~&
\,\,  \epsilon^4~~ &\,\, \epsilon^3
\\
\,\, \epsilon^4 ~~ &\,\,\epsilon^3~~&\,\, \epsilon^2~
\end{array}\right)\overline{H}\epsilon^a }~,
\end{array}  \!\!~
\label{15-5}
\eeq

\begin{table}
\caption{Transformation properties of matter superfields 
under ${\cal U}(1)$ symmetry}
\label{t:fer}
$$\begin{array}{|c|c|c|c|c|c|c|c|c|}
\hline
& & & & & & & & \\
   & 10_3& 10_2&
10_1&\bar 5_i &
15_2  & \overline{15}_2 &
15_1  &\overline{15}_1
 \\
& & & & & & & & \\
\hline
& & & & & & & & \\
{\cal U}(1)&r  &r-1 &r-3 &
-(3r+a+3-i)  &r-2  &1-r  &
r-3 &3-r  \\
& & & & & & & & \\
\hline
\end{array}$$
\end{table}

\begin{equation}
\begin{array}{cc}
 & {\begin{array}{ccc}
10_1&\,\,10_2&\,\,10_3~~~
\end{array}}\\ \vspace{2mm}
\begin{array}{c}
\overline{15}_1\\ \overline{15}_2

\end{array}\!\!\!\!\! &{\left(\begin{array}{ccc}
\,\, 1~~&
\,\,  0~~ &\,\, 0
\\
\,\, \epsilon^2 ~~ &\,\,1~~&\,\, 0~
\end{array}\right)\Sigma }~,
\end{array}  \!\!~
\begin{array}{cc}
 & {\begin{array}{cc}
15_1&\,\,
15_2~~~~~~~~~
\end{array}}\\ \vspace{2mm}
\begin{array}{c}
\overline{15}_1 \\ \overline{15}_2

\end{array}\!\!\!\!\! &{\left(\begin{array}{ccc}
\,\, 1~~
 &\,\,0
\\
\,\, \epsilon^2~~
&\,\,\epsilon
\end{array}\right)(\Sigma+S).
}
\end{array}
\label{15s}
\end{equation}

Noting that in terms of $SU(3)_c\times SU(2)_W $, 
$15=(3, 2)+(6, 1)+(1, 3)$, we may conclude that the couplings involving
$15, \overline{15}$ do not affect $e^c$ and $l$ states from $10$ and
$\bar 5$
respectively (they can only affect the $q$ states). The lepton mass matrix
will coincide with (\ref{10-5}),
from which we have:

\beq
\lambda_{\tau }\sim \epsilon^a~,~~~
\lambda_e:\lambda_{\mu }:\lambda_{\tau }\sim 
\epsilon^5:\epsilon^2:1~,
\label{lep}
\eeq 
where $a=0, 1, 2$ determines the value of 
$\tan \beta $($\sim \frac{m_t}{m_b}\epsilon^a$).

Turning to the quark sector, from (\ref{15s}) we see that $10_3$-plet
also 
is not affected, while $q_{10_1}, q_{10_2}$ will be mixed with 
$q_{15_1}, q_{15_2}$. Analyzing (\ref{15s}), one can easily verify that
for the `light' $q_i$ states we will have:

$$
(10_1,~15_1)\stackrel {\supset}{_\sim }q_1~,~~~15_2 \supset q_2~,
$$
\beq
10_2 \stackrel {\supset}{_\sim} \epsilon q_2~,~~~~~~~~10_3 \supset q_3~.
\label{weights}
\eeq
From (\ref{weights}), (\ref{10-5}) and (\ref{15-5}),  we find
the  down quark mass matrix to be

\begin{equation}
\begin{array}{ccc}
 & {\begin{array}{ccc}
\hspace{-5mm}~d^c_1~& \,\,d^c_2~~ & \,\,d^c_3~~~

\end{array}}\\ \vspace{2mm}
\begin{array}{c}
q_1 \\ q_2 \\ q_3
 \end{array}\!\!\!\!\! &{\left(\begin{array}{ccc}
\,\,\epsilon^5 ~~ &\,\,\epsilon^4~~ &
\,\,\epsilon^3
\\
\,\,\epsilon^4 ~~ &\,\,\epsilon^3~~ &
\,\,\epsilon^2
 \\
\,\,\epsilon^2 ~~ &\,\,\epsilon~~ &\,\,1
\end{array}\right)\epsilon^ah_d }~,
\end{array}  \!\!  ~~~~~
\label{down}  
\end{equation}
from which

\beq
\lambda_b\sim \epsilon^a~,~~~
\lambda_d:\lambda_s:\lambda_b\sim 
\epsilon^5:\epsilon^3:1~.
\label{masdown}
\eeq

From (\ref{10-5}), (\ref{lep}), (\ref{down}), (\ref{masdown}), and
taking into account (\ref{weights}), we obtain

\beq
\lambda_b=\lambda_{\tau }\left(1+{\cal O}(\epsilon^2)\right)\sim 
\epsilon^a~,
\label{btau}
\eeq
while, for Yukawas of the second generation,

\beq
\lambda_s \sim \epsilon \lambda_{\mu }\simeq 
\frac{1}{5}\lambda_{\mu }~.
\label{smu} 
\eeq
Assuming that $\lambda_d \sim 2\lambda_e $, from (\ref{lep})
(\ref{masdown}) and (\ref{smu}) we will have

\beq
\frac{\lambda_s}{\lambda_d}\sim 
\frac{1}{10}\frac{\lambda_{\mu}}{\lambda_e}\simeq 20~.
\label{sd}
\eeq

For up-type quarks, from (\ref{10-10}), taking into
account (\ref{weights}), we  obtain

\begin{equation}
\begin{array}{ccc}
 & {\begin{array}{ccc}
\hspace{-5mm}~u^c_1~~& \,\,u^c_2~ & \,\,u^c_3

\end{array}}\\ \vspace{2mm}
\begin{array}{c}
q_1 \\ q_2 \\ q_3
 \end{array}\!\!\!\!\! &{\left(\begin{array}{ccc}
\,\,\epsilon^6 ~~ &\,\,\epsilon^4~~ &
\,\,\epsilon^3
\\
\,\,\epsilon^4 ~~  &\,\,\epsilon^3~~ &
\,\,\epsilon^2
 \\
\,\,\epsilon^3~~ &\,\,\epsilon~~ &\,\,1
\end{array}\right)h_u }~,
\end{array}  \!\!  ~~~~~
\label{up}  
\end{equation}
from which we obtained the desired Yukawa couplings

\beq
\lambda_t\sim 1~,~~~\lambda_u:\lambda_c:\lambda_t\sim
\epsilon^6:\epsilon^3:1~.
\label{masup}
\eeq

From (\ref{down}) and (\ref{up}), for the CKM matrix elements we find

\beq
V_{us}\sim \epsilon~,~~~V_{cb}\sim \epsilon^2~,~~~
V_{ub}\sim \epsilon^3~,
\label{ckm}
\eeq
in good agreement with observations.

To conclude, we see that with the help of ${\cal U}(1)$ flavor symmetry
and $\overline{15}+15$-plets, in addition to  the desirable 
hierarchies of charged fermion  masses
and CKM mixing angles, we can also get reasonable [(\ref{btau}), 
(\ref{smu}), (\ref{sd})] asymptotic relations.

\section{Value of $\alpha_s(M_Z)$}

By analyzing the spectra of decoupled heavy states, from (\ref{15s}) we
can verify
that the masses of the states $(\bar 6,1)+(1,\bar 3)+(6,1)+(1,3)$
(from $\overline{15}_2+15_2$ respectively) are below the GUT scale and
equal to $M_S\simeq M_G\epsilon $. Indeed, these states will change the
running of the gauge couplings  above the $M_S$ scale and, as we
will
see, this opens up the  possibility to obtain a reduced value for
$\alpha_s(M_Z)$
\footnote{For alternative mechanisms of achieving this see \cite{ilo}.}.

The solutions of the three renormalization-group (RG) equations are
\cite{lan} 

\beq
\alpha_G^{-1}=\alpha_a^{-1}-\frac{b_a}{2\pi }\ln \frac{M_G}{M_Z}
-\frac{b_a'}{2\pi }\ln \frac{M_G}{M_S}+\Delta_a+\delta_a~,
\label{rg}
\eeq
where $\alpha_G$ is the gauge coupling at the GUT scale,
$\alpha_a$
denote the  gauge couplings at $M_Z$ scale
($\alpha_{1,2,3}$ are the gauge couplings of $U(1)_Y$, $SU(2)_W$ and
$SU(3)_c$ respectively), while $b_a$, $b_a'$ are given by

\beq
(b_1,~b_2,~b_3)=(\frac{33}{5},~1,~-3)~,~~~~
(b_1',~b_2',~b_3')=(\frac{34}{5},~4,~5)~.
\label{bs}
\eeq
The $\Delta_a $ include all possible SUSY and heavy threshold corrections,
and
contributions from the two loop effects of MSSU(5). 
$\delta_a$ denote the difference of gauge coupling running between
MSSU(5) and present model from $M_S$ up to $M_G$ in two loop
approximation,

\beq
\delta_a=\frac{1}{4\pi }
\left(\frac{b_{ab}+b_{ab}'}{b_b+b_b'}\ln \frac{\alpha_b(M_S)}{\alpha_G}-
\frac{b_{ab}}{b_b}\ln \frac{\alpha_b(M_S)}{\alpha_G^0}
\right)~, 
\label{2loop}
\eeq
where 

\begin{equation}
\begin{array}{ccc}
 & {\begin{array}{ccc}
&\,\,&\,\,~~~
\end{array}}\\ \vspace{2mm}
\begin{array}{c}
\\ 

\end{array}\!\!\!\!\! &b_{ab}={\left(\begin{array}{ccc}
\,\, \frac{199}{25}~~&
\,\,  \frac{27}{5}~~ &\,\, \frac{88}{5}
\\
\,\, \frac{9}{5} ~~ &\,\,25~~&\,\, 24~
\\
\,\,\frac{11}{5}~~ &\,\,9~~ &\,\,14
\end{array}\right)}~,
\end{array}  \!\!~
\begin{array}{ccc}
 & {\begin{array}{ccc}
&\,\,
~~~~~~
\end{array}}\\ \vspace{2mm}
\begin{array}{c}
 \\ 

\end{array}\!\!\!\!\! &b_{ab}'={\left(\begin{array}{ccc}
\,\, \frac{904}{75} ~~ &\,\,\frac{144}{5}~~&\,\,\frac{128}{3}~
\\
\,\,\frac{144}{15}~~ &\,\,24~~&\,\,0~
\\
\,\, \frac{16}{3}~~ &\,\,0~~&\,\, \frac{128}{3}~
\end{array}\right)
}
\end{array}
\label{bb}
\end{equation}
and the appropriate couplings in (\ref{2loop}) are calculated in
one loop approximation. $\alpha_G^0$ is the gauge coupling at $M_G$ in 
MSSU(5).

From (\ref{rg}), taking into account (\ref{bs}), 
one finds 

\beq
\alpha_s^{-1}=\left(\alpha_s^{-1} \right)^{0}+
\frac{3}{2\pi }\ln \frac{M_G}{M_S}+\delta ~,
\eeq
where
$\left(\alpha_s^{-1} \right)^{0}=\frac{1}{7}
\left(12\alpha_w^{-1}-5\alpha_Y^{-1}\right)+
\frac{1}{7}\left(12\Delta_2-5\Delta_1-7\Delta_3\right)
$ corresponds to the value of $\alpha_s$ obtained in MSSU(5)
case, and $\delta=\frac{1}{7}(12\delta_2-5\delta_1-7\delta_3)$. 
Using the result $\left(\alpha_s^{-1} \right)^{0}=1/0.126$ 
\cite{lan}, 
and taking $M_S/M_G\simeq \epsilon \simeq 0.2$, (neglecting $\delta $ for
the time being),
we obtain $\alpha_s\simeq 0.115$, in  good agreement with experimental
data \cite{dat}.
Taking into account (\ref{rg}) and (\ref{bb}), from (\ref{2loop}) we
obtain $\delta=-0.015$, thus
leaving the value of $\alpha_s$ unchanged as expected.

%
%

\section{Proton Decay}

From (\ref{10-5}) and (\ref{15-5}), taking into account 
(\ref{weights}), we see
that $ql\bar T$ type couplings in the family space have the
same hierarchical structure as the down quark mass matrix (\ref{down}).
As far as $qqT$ operators are concerned, from (\ref{10-10}), 
(\ref{weights}) one obtains,

\begin{equation}
\begin{array}{ccc}
 & {\begin{array}{ccc}
\hspace{-5mm}~q_1~& \,\,q_2 & \,\,~q_3

\end{array}}\\ \vspace{2mm}
\begin{array}{c}
q_1 \\ q_2 \\ q_3
 \end{array}\!\!\!\!\! &{\left(\begin{array}{ccc}
\,\,\epsilon^6 ~~ &\,\,\epsilon^5~~ &
\,\,\epsilon^3
\\
\,\,\epsilon^5 ~~  &\,\,\epsilon^4~~ &
\,\,\epsilon^2
 \\
\,\,\epsilon^3~~ &\,\,\epsilon^2~~ &\,\,1
\end{array}\right)T }~,
\end{array}  \!\!  ~~~~~
\label{qqT}  
\end{equation}
from which we see that the appropriate couplings are suppressed by a 
factor
$\epsilon(\sim 1/5)$ compared to the up type quark mass matrix
(\ref{up}). From (\ref{rg}), we find that 
$M_G=\left(\frac{M_S}{M_G}\right)^{1/2}M_G^0\simeq M_G^0/\sqrt{5}$, where
$M_G^0$ is the GUT scale in MSSU(5). From all this we may conclude that
the proton life time in our model will be
$\tau_p\sim 5\cdot \tau_p^0$ (that is, a factor $5$ larger than in
MSSU(5)).
For further suppression of nucleon decay, the mass scale $M_S$ should be
reduced. However, this would ruin the gauge coupling unification
unless some additional mechanism (for retaining unification) is
applied. Such a program can be
successfully realized in extended $SU(5+N)$ GUTs \cite{suN}.

\section{Neutrino Oscillations}

Turning to the neutrino sector, for accommodating the recent solar and
atmospheric Superkamiokande data  (see \cite{sol}, \cite{atm}
respectively), we will invoke the
mechanism suggested in
refs. \cite{numssm, bimax}. The
atmospheric anomaly is explained through maximal $\nu_{\mu }-\nu_{\tau }$
mixings which is  achieved through quasi-degenerate massive $\nu_{\mu}$,
$\nu_{\tau }$ states. Since these states are too heavy to explain the
solar neutrino data, we are led introduce
a sterile neutrino state $\nu_s$. The solar
neutrino anomaly is resolved via the small angle $\nu_e-\nu_s$ MSW
oscillations.

Together with $\nu_s$ state we introduce two heavy right handed states
${\cal N}_{2,3}$. Choosing the ${\cal U}(1)$ charges of these states 
to be

\beq
Q_{{\cal N}_2}=-\frac{1}{2}~,~~~~
Q_{{\cal N}_3}=\frac{1}{2}~,~~~~Q_{\nu_s }=-\frac{41}{2}~,
\label{charges}
\eeq
and in Table (\ref{t:fer}) taking

\beq
r=-\frac{a}{5}-\frac{1}{10}~,
\label{chs}
\eeq  
the relevant couplings are
(these singlet states do not transform under the $Z_2$ ${\cal R}$
symmetry):

\begin{equation}
\begin{array}{cc}
 & {\begin{array}{cc}
~~{\cal N}_2&\,\,{\cal N}_3~~~
\end{array}}\\ \vspace{2mm}
\begin{array}{c}
\bar 5_1\\\bar 5_2\\ \bar 5_3 \\
 
\end{array}\!\!\!\!\! &{\left(\begin{array}{ccc}
\, \epsilon^2~ &
\,\,~\epsilon   
\\
\, \epsilon~ &
\,\,~1
\\
\, 1 &\,\,~0
\end{array}\right)H }
\end{array}  \!\!~,~~~~~~~~~~~
\begin{array}{cc}
 & {\begin{array}{cc}
~{\cal N}_2&\,\,
{\cal N}_3~~~~
\end{array}}\\ \vspace{2mm}
\begin{array}{c}
{\cal N}_2 \\ {\cal N}_3

\end{array}\!\!\!\!\! &{\left(\begin{array}{ccc}
\, \epsilon~~
 &\,\,1
\\    
\, 1~~
&\,\,0
\end{array}\right)\rho S}
\end{array} ~,~~
\label{fNNN}
\end{equation}

\begin{equation}
W_{\nu s}=\epsilon^{20}
\left(\bar 5_3+\epsilon \bar 5_2+\epsilon^2\bar 5_1 \right)
\nu_sH+
S\epsilon^{41}\nu_s^2~,
\label{wsol}
\end{equation}   
where $\rho $ is a dimensionless coupling.
Integration of ${\cal N}_{2,3}$ states leads to the mass matrix 
for the `light' neutrinos:

\begin{equation}
\begin{array}{cccc}
 & {\begin{array}{cccc}
\hspace{-5mm}~~~~\nu_s & \,\,~~~~~\nu_e~~~  &
\,\,~\nu_{\mu }~~& \,\,~\nu_{\tau }
\end{array}}\\ \vspace{2mm}
m_{\nu }= \begin{array}{c}
\nu_s \\ \nu_e \\ \nu_{\mu } \\ \nu_{\tau }
 \end{array}\!\!\!\!\! &{\left(\begin{array}{cccc}
\,\,m_{\nu_s}  &\,\,~~m'\epsilon^2 &
\,\,~m'\epsilon &~m' 
\\
\,\,m'\epsilon^2 &\,\,~~m\epsilon^3  &
\,\,~m\epsilon^2&m   
 \\
\,\,m'\epsilon &\,\,~~m\epsilon^2 &
\,\,~m\epsilon & m
\\
\,\,m' &\,\,~~ m\epsilon &\,\,~m &~~0
\end{array}\right) }~,
\end{array}  \!\!  ~~~~~
\label{matst1}
\end{equation}
where we have defined:

\beq
m\equiv \frac{h_u^2}{\rho M_P\epsilon_G}~,~~~
m'\equiv \epsilon^{20}h_u~,~~~
m_{\nu_s }\equiv M_P\epsilon_G\epsilon^{41}~.
\label{agn}
\eeq
Taking  $\rho \sim 2\cdot 10^{-2}$, $\epsilon =0.2-0.22$, from
(\ref{agn}) we have

$$
m\simeq 6.3\cdot 10^{-2}{\rm eV}~,
$$
$$
m_{\nu_s}=(5\cdot 10^{-4}-3\cdot 10^{-2}){\rm eV}~,   
$$
\beq
m'=(1.8\cdot 10^{-3}-1.2\cdot 10^{-2}){\rm eV}.
\label{saz}
\eeq

Note that the sterile neutrino is kept light (see (\ref{wsol}),
(\ref{saz}))
by the ${\cal U}(1)$ symmetry \cite{chun, numssm}.
Taking
\beq
m=6.3\cdot 10^{-2}{\rm eV}~,~~~
m'=1.8\cdot 10^{-3}{\rm eV}~,~~~
m_{\nu_s}=2\cdot 10^{-3}{\rm eV}~
\label{ranges}
\eeq
from (\ref{ranges}) and (\ref{matst1}), we have for the atmospheric
neutrino
oscillation parameters 
  
$$
\Delta m_{23}^2=2m^2\epsilon \simeq 2\cdot 10^{-3}~{\rm eV}^2~,
$$
\begin{equation}
\sin^2 2\theta_{\mu \tau } =1-{\cal O}(\epsilon^2)~.
\label{atm}
\end{equation}
The solar neutrino oscillation parameters are given by

$$
\Delta m_{\nu_e \nu_s}^2\simeq m_{\nu_s}^2
\sim  4\cdot 10^{-6}~{\rm eV}^2~,
$$
\begin{equation}
\sin^2 2\theta_{es}\simeq
4\left( \frac{m'\epsilon^2}{m_{\nu_s }}\right)^2
\sim 5\cdot  10^{-3} ~.
\label{sol}
\end{equation}

We see that the ${\cal U}(1)$ flavor symmetry helps provide a natural
explanation of the solar and atmospheric experimental data.
Note that $a$ is still undetermined, and therefore the magnitude of
$\tan \beta $ is not fixed in our model.

\section{Automatic Matter Parity}

Let us conclude by considering all possible `matter' parity violating
operators:

$$
\bar 5_iH~,~~~~10_i\bar H(\Sigma \bar H)~,~~~~(\Sigma+S)15_i\bar H\bar
H~,~~
$$
$$
(\Sigma+S)\overline{15}_iHH~,~~~(\Sigma+S)10_i\bar 5_j\bar 5_k~,~~~
10_i10_j10_k\bar H~,
$$
\beq
(\Sigma+S)15_i\bar 5_j\bar 5_k~,~~~~~~\Sigma^2 15_i15_j15_k\bar H~,
~\dots
\label{matpar}
\eeq
From Table (\ref{t:fer}), taking into account (\ref{chs}), we observe
that the terms in (\ref{matpar}) all have non-integer ${\cal
U}(1)$ charges,
and consequently are forbidden to `all orders' in powers of $X$.
Therefore, thanks to ${\cal U}(1)$
flavor symmetry, the model has automatic matter parity. 

\section{Conclusion}

In conclusion, we note that the mechanisms discussed here for
resolving the various puzzles in $SU(5)$ can be successfully generalized
to $SU(5+N)$ GUTs \cite{suN}.
In this paper we have not addressed the gauge hierarchy problem whose
resolution in $SU(5)$ requires additional 'scalar' multiplets belonging
to $50+\overline{50}+75$. In such a scenario the Higgs doublets
remain 'massless', while the color triplets obtain masses by mixing with
the triplets in $50, \overline{50}$. On the other hand, in order to
retain perturbative gauge couplings
up to $M_P$, the masses of $50, \overline{50}$ states should exceed
$M_G$, which means that
the ordinary color triplets (from $H, \bar H$) will lie below $M_G$.
This would further destabilize the proton, and possibly disrupt unification
of the gauge couplings. To avoid this, one could either consider more
complicated $SU(5)$ scenarios \cite{sce} with extended scalar sector,
or extended $SU(5+N)$ GUTs \cite{suN}. In the
latter case, for instance, a $SU(6)$ model has been discussed in which
the MSSM Higgs doublets  are pseudo-Goldstone bosons, 
the proton lifetime is $\sim 10^2 \tau_p^{SU(5)}$,
and neutrino oscillations involve bi-maximal mixings.


\begin{thebibliography}{99}

\bibitem{dim}
S. Dimopoulos and H. Georgi, Nucl. Phys. B 193 (1981) 150;
N. Sakai, Z. Phys. C 11 (1982) 153.

\bibitem{lan}
P. Langacker, N. Polonsky, Phys. Rev. D 52 (1995) 52;
J. Bagger, K. Matchev, D. Pierce, Phys. Lett. B 348 (1995) 443.

\bibitem{dat}
Particle Data Group, Phys. Rev. D 54 (1996) 1.

\bibitem{bimax}
Q. Shafi, Z. Tavartkiladze, hep-ph/9901243, Phys. Lett. B 451 (1999) 129.

\bibitem{fro} 
C.D. Frogatt, H.B. Nielsen, Nucl. Phys. B 147 (1979) 277.

\bibitem{GJ}
H. Gorgi, C. Jarlskog, Phys. Lett. B 88 (1979) 279;
J. Harvey, P. Ramond, D. Reiss, Phys. Lett. B 92 (1980) 309;
S. Dimopoulos, L.J. Hall, S. Raby, Phys. Rev. Lett. 68 (1992) 1984,
 Phys. Rev. D 45 (1992) 4192;
V. Barger et. al., Phys. Rev. Lett. 68 (1992) 3394;
H. Arason et. al., Phys. Rev. D 47 (1993) 232.


\bibitem{gsh}
M. Green and J. Schwarz, Phys. Lett. B 149 (1984) 117.


\bibitem{fi}
 E. Witten, Nucl. Phys. B 188 (1981) 513;
W. Fischler et al., Rhys. Rev. Lett. 47 (1981) 657.

\bibitem{xi}
 M. Dine, N. Seiberg and E. Witten, Nucl. Phys. B 289
(1987) 584;
J. Atick, L. Dixon and A. Sen, Nucl. Phys. B 292 (1987) 109; M. Dine,
I. Ichinose and N. Seiberg, Nucl. Phys. B 293 (1987) 253.

\bibitem{gia}
G. Dvali and A. Pomarol, Phys. Rev. Lett. 77 (1996) 3738;
P. Binetruy and E. Dudas, Phys. Lett. B 389 (1996) 503.

\bibitem{u1}
L. Iba\~nez and G.G. Ross, Phys. Lett. B 332 (1994) 100; 
P. Binetruy and P. Ramond, Phys. Lett. B 350 (1995) 49; 
V. Jain and R. Shrock, Phys. Lett. B 352 (1995) 83; 
E. Dudas, S. Pokorski and C. Savoy,  
Phys. Lett. B 369 (1995) 255; 
P. Binetruy, S. Lavignac and P. Ramond, Nucl. Phys.
B 447 (1996) 353.
Z. Berezhiani, Z. Tavartkiladze, Phys. Lett. B 396 (1997) 150;
B 409 (1997) 220;
A. Nelson, D. Wright, Phys. Rev. D 56 (1997) 1598;
N. Irges, S. Lavignac, P. Ramond, Phys. Rev. D 58 (1998) 035003;
J. Elwood, N. Irges, P. Ramond, Phys. Rev. Lett. 81 (1998) 5064.

\bibitem{ilo}
See second ref. of \cite{lan} and also
I. Gogoladze, hep-ph/9612365; 
J. Chkareuli, I. Gogoladze, Phys. Rev. D 58 (1998) 055011.


\bibitem{suN}
Q. Shafi, Z. Tavartkiladze, hep-ph/9905202.


\bibitem{sol}
 J.N. Bahcall, P.I. Krastev, A.Yu. Smirnov,
Phys. Rev. D 58 (1998) 096016; See also references therein.


\bibitem{atm}
T. Kajita, hep-ex/9810001.


\bibitem{numssm}
Q. Shafi, Z. Tavartkiladze, hep-ph/9811463, Phys. Lett. B 448 (1999) 46.

\bibitem{chun}
E.J. Chun, A.S. Joshipura, A.Yu. Smirnov,
Phys. Lett. B 357 (1995) 608; Phys. Rev. D 54 (1996) 4654;   
Q. Shafi and Z. Tavartkiladze, hep-ph/9807502, hep-ph/9811282,
to appear in Nucl. Phys. B.
   
\bibitem{sce}
J. Hisano, T. Moroi, K. Tobe, T. Yanagida, Phys. Lett. B 342 (1995) 138;
J. Hisano, Prog. Theor. Phys. Suppl. 123 (1996) 301;
See also Z. Berezhiani, Z. Tavartkiladze in \cite{u1}. 





\end{thebibliography}
\end{document}